\documentclass[prb,a4,twocolumn,showpacs,amsmath,amssymb]{revtex4-1}

\usepackage{graphicx,subfigure,epsfig}
\usepackage{color}
\usepackage{soul}
\usepackage{float}
\usepackage{amsmath}
\usepackage{amssymb}
\usepackage{ifthen}
\usepackage{alltt}
\usepackage{fancyhdr}
\usepackage{verbatim}
\usepackage{moreverb}
\usepackage{colortbl}
\usepackage{multirow}
\usepackage{bigstrut}
\usepackage{wrapfig}
\usepackage{enumerate}
\usepackage{rotating}
\usepackage{epstopdf}
\def\e{\begin{equation}}
\def\f{\end{equation}}

\def\r#1{(\ref{eq:#1})}

\begin{document}

\title{Magnetic hyperbolic metamaterial of polaritonic nanowires}
\author{M.~S.~Mirmoosa$^1$, S.~Yu.~Kosulnikov$^{1,2}$ and C.~R.~Simovski$^{1,2}$}
\affiliation{$^1$Department of Radio Science and Engineering, School of Electrical Engineering, Aalto University, P.O.~Box 13000, FI-00076 AALTO, Finland\\
$^2$Department of Nanophotonics and Metamaterials, ITMO University, 197043, St.~Petersburg, Russia}
\date{\today }


\begin{abstract}
We show that the axial component of the magnetic permeability tensor is resonant for a wire medium consisting of high-index epsilon-positive nanowires, and its real part
changes the sign at a certain frequency. At this frequency the medium experiences the topological transition from the hyperbolic to the elliptic type of dispersion.
We show that the transition regime is characterized by extremely strong dependence of the permeability on the
wave vector. This implies very high density of electromagnetic states that results in the filamentary pattern
and noticeable Purcell factor for a transversely oriented magnetic dipole.
\end{abstract}

\maketitle


\section{introduction}
Simple wire media, defined as optically dense arrays of parallel metal wires in a host dielectric material, have been in a considerable attention since electromagnetic metamaterials have been introduced.
Many applications of such media have been suggested from radio frequencies where the period of wire media is on the millimeter scale, to optical frequencies where this period is submicron. Among them one can mention biological sensing \cite{kabashin}, subwavelength imaging and endoscopy with image magnification \cite{belov1,casse,wan}, enhancement of quantum emitters, thermal sources, and classical dipole radiators \cite{poddubny1,simovski,mirmoosa2}, enhancement of radiative heat transfer \cite{simovski2,mirmoosa1}, etc. More details can be found in overview \cite{simovski1}. In all these works, electromagnetic properties of wire media were described in a condensed form -- via an effective permittivity tensor:
\e
\overline{\overline{\epsilon}}=\left(\begin{array}{ccc}
\varepsilon_{\perp} & 0 & 0 \\
0 &\varepsilon_{\perp} & 0 \\
0& 0 & \varepsilon_{\parallel} \end{array} \right).
\label{e1}\f
Here, $\varepsilon_\perp$ and $\varepsilon_\parallel$ are the perpendicular (transversal) and parallel (axial) components with respect to the wires axis. Therefore, standard wire media are uniaxially anisotropic materials (though not usual ones because they possess spatial dispersion -- dependence of  $\overline{\overline{\epsilon}}$ on the wave vector ${\bf k}$ \cite{mario2,poddubny2}.

At infrared frequencies, the relative dielectric constant of the metal nanowires is highly negative. If the fraction ratio of the metal in the effective medium is low the perpendicular component $\varepsilon_\perp$ of the permittivity is approximately equal to the permittivity of the host medium, whereas the parallel component $\varepsilon_\parallel$ has negative sign \cite{mario2,simovski1}. For the dispersion, it implies that the transverse magnetic (TM) polarized wave has an open dispersion surface similar to a hyperboloid \cite{simovski1,poddubny2}. This dispersion surface results in the high density of electromagnetic states and enhancement of a subwavelength electric dipole located in the medium orthogonally to the wires \cite{poddubny1,poddubny2}. The radiation of a magnetic dipole can be also enhanced but this effect is lower because such the dipole produces mainly TE-waves whose electric field is orthogonal to nanowires. These transversely electric (TE) polarized waves weakly interact with the wire medium and their dispersion surfaces are rather similar to spheres like dispersion surfaces of free space. Notice, that for metal nanowires the effective permeability tensor ($\overline{\overline{\mu}}$) is practically equal to unity \cite{simovski1}.

Recently, in Ref.~\cite{mario1} one studied an anisotropic medium with artificial magnetism. This medium is dual to the dielectric hyperbolic metamaterial described by the indefinite tensor in Eq.~\r{e1}. In this magnetic metamaterial the effective permittivity tensor is definite, positive and not resonant. Therefore, the dispersion surfaces of TM-polarized waves in this medium are similar to the spheres, whereas
the dispersion surfaces of TE-waves are either hyperbolic or elliptic depending on frequency. However, medium in Ref.~\cite{mario1} is not a wire medium. It is a racemic array of metal helices with both left-handedness and right-handedness. Such media, in accordance to initial works \cite{spiral,spiral1} where they were introduced, should be referred as spiral media. Spiral medium of \cite{mario1} operates at microwave frequencies, where metals are close to perfect conductors. In the present paper, we study wire media at optical frequencies.

\begin{figure}[t!]\centering
\includegraphics[width=6.5cm]{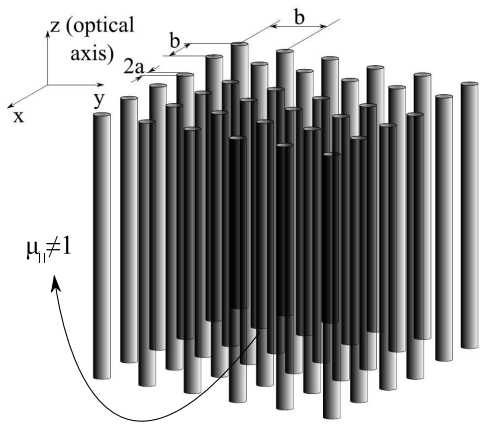}
\caption{A medium of long parallel wires which are set in a square lattice. The wires are made of polaritonic materials whose relative dielectric constant is much larger than unity ($\varepsilon_{\rm{r}}\gg1$). Here, $a$ and $b$ are the wires radius and lattice constant, respectively. For simplicity, the host medium is free space.}
\label{fig:medium}
\end{figure}

Below, we theoretically demonstrate that a simple wire medium (see Fig.~\ref{fig:medium}) may have in the infrared range similar electromagnetic properties to those manifested by \cite{mario1} exhibited at microwaves, however, there are also significant peculiarities which share our wire medium out from all hyperbolic metamaterials. Nanowires of our wire medium should be made of high-index epsilon-positive materials, such as lithium tantalate, silicon carbide or hexagonal boron nitride, and some other polaritonic materials in the range 15--30 THz.
The resonance of the axial component of the effective permeability tensor arises because the dynamic magnetic polarizability of the high-index dielectric wire is resonant, and origins from the Mie resonance of a single nanowire. We show that this axial permeability changes its sign at a certain frequency and in a narrow frequency range around this transition is a very sharp function of the wave vector.
The corresponding topological transition is related to the resonant enhancement of radiation of a subwavelength dipole located inside the medium. It is an expected effect which is similar to that studied in \cite{mirmoosa2}.
However, in the present case it holds for a magnetic dipole oriented transversely to nanowires. The sharp dependence of the permeability on the wave vector results in the unusual property of our wire medium --
the radiation of the internal magnetic dipole is concentrated at the line which passes through the dipole center in parallel to the optical axis. This property is difficult to detect from the dispersion contours (sections of the dispersion surfaces). In the present case these contours for TE-waves do not differ very much from analogous contours for TM-waves we have studied in \cite{mirmoosa1}. Moreover, these contours were analyzed for TE-waves in the spiral medium \cite{mario1}. Therefore, we do not spend the volume of this paper to the demonstration and analysis of the dispersion contours in the region of topological transition and fully concentrate on two effects: pattern of the internal source and its Purcell factor.

The paper is organized as follows: In Section~\ref{sec:ep}, we obtain analytical expressions for the magnetic polarizability of a high-index dielectric cylinder and the effective permeability of the corresponding wire medium. In Section~\ref{sec:pf}, we discuss the implications of the resonant and non-local behavior of this  permeability and show the simulated results for the radiation of an embedded magnetic dipole. In Section~\ref{sec:con}, we present the brief summary of the work.


\section{Effective permeability of the wire medium}

\label{sec:ep}
\begin{figure}[t!]\centering
\includegraphics[width=6.5cm]{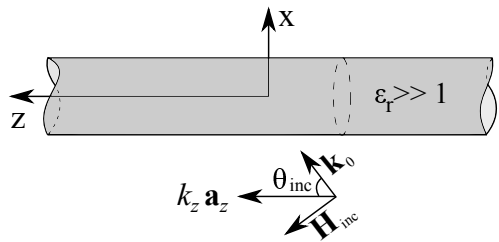}
\label{fig:cylinder}
\caption{(a)--An infinite dielectric cylinder illuminated by a TE-polarized plane wave.}
\end{figure}

Let a plane wave with electric field polarized along $\mathbf{y}_0$ illuminate a dielectric cylinder, as it is shown in Fig.~\ref{fig:cylinder}. The incident wave has perpendicular polarization and its magnetic field along $\mathbf{z}_0$ can be written as
\e
H_{z_{\rm{inc}}}=H_0e^{-j(k_xx+k_zz)},
\label{H}\f
where $H_0$ is the magnitude, $k_z$ and $k_x$ are the wave vector components in the free space. Due to the cylindrical geometry, it is proper to expand the field into cylindrical harmonics. Therefore,
\begin{equation}
H_{z_{\rm{inc}}}=H_0\sum_m\left[(-j)^nJ_m(h_0R)e^{jm\phi}\right]e^{-jk_zz},
\end{equation}
in which $J_m(x)$ is the Bessel function (first kind, order $m$), $h_0=\sqrt{k_0^2-k_z^2}$ ($k_0$ is the free-space wave number) and $(R, \phi, z)$ are the components of the cylindrical coordinate system.
Taking into account the scattering and penetration of the wave into the dielectric cylinder, the total electromagnetic fields outside and inside the cylinder can be described by
\begin{equation}
\begin{split}
&E_{z_{\rm{out}}}=\sum_m{C_mH_m^{(2)}(h_0R)e^{jm\phi}},\cr
&H_{z_{\rm{out}}}=\sum_m \left(D_mH_m^{(2)}(h_0R)+H_0(-j)^mJ_m(h_0R)\right)e^{jm\phi},
\end{split}
\label{field1}
\end{equation}
and
\begin{equation}
\begin{split}
&E_{z_{\rm{in}}}=\sum_m{A_mJ_m(hR)e^{jm\phi}},\cr
&H_{z_{\rm{in}}}=\sum_m{B_mJ_m(hR)e^{jm\phi}},
\end{split}
\label{field2}
\end{equation}
where $h=\sqrt{k_0^2\varepsilon_{\rm{r}}-k_z^2}$ ($\varepsilon_{\rm{r}}$ denotes the relative dielectric constant of the cylinder) and $H^{(2)}_m(x)$ represents the Hankel function (second kind, order $m$). The other components of the electromagnetic fields ($E_R$, $H_R$, $E_\phi$ and $H_\phi$) can be readily derived by solving the Maxwell's equations. In Eqs.~(\ref{field1}) and (\ref{field2}), the unknown coefficients ($A_m$, $B_m$, $C_m$ and $D_m$) are related to each other through boundary conditions, i.e.~the tangential components of the electric and magnetic fields should be continuous at the radius of the cylinder ($R=a$). After determining the fields inside the cylinder, we can obtain the polarization current that is equal to $\mathbf{J}_{\rm{p}}=j\omega\varepsilon_0(\varepsilon_{\rm{r}}-1)\mathbf{E}_{\rm{in}}$. On the other hand, we know that the magnetic moment per unit length of our cylinder is described by
\begin{equation}
\mathbf{m}=\frac{1}{2}\mu_0\int_S\mathbf{r}\times\mathbf{J}_{\rm{p}}dS,
\end{equation}
where $\mathbf{r}=\mathbf{R}+z\mathbf{a}_z$ is the distance vector from the origin to the current element. Therefore, the $\mathbf{z}$-component of the magnetic moment can be obtained as
\begin{equation}
m_z=\frac{1}{2}j\omega\mu_0\varepsilon_0(\varepsilon_{\rm{r}}-1)\int_SR^2E_{\phi_{\rm{in}}}dRd\phi,
\label{moment}
\end{equation}
in which the $\phi$-component of the electric field inside the cylinder is as follows:
\begin{equation}
\begin{split}
E_{\phi_{\rm{in}}}=&\frac{j\omega\mu_0}{h}\times\cr
&\sum_m\left(A_m\frac{mk_z}{j\omega\mu_0hR}J_m(hR)+B_mJ'_m(hR)\right)e^{jm\phi}.
\end{split}
\label{Ephi}
\end{equation}
The function $J'_m(x)$ is the derivative of $J_m(x)$. From the Eqs.~(\ref{moment}) and (\ref{Ephi}), we can see that the only cylindrical harmonic which is responsible for non-zero magnetic moment per unit length produced by the magnetic field of the incident wave is $m=0$. Then, based on Eq.~(\ref{Ephi}), we should only calculate the coefficient $B_0$ -- $A_0$ gives the zero contribution into  Eqs.~(\ref{moment}). From imposing the boundary conditions to determine the unknown coefficients, we can achieve
\e
B_0=\frac{j2h}{\pi h_0a\left[h_0J'_0(ha)H^{(2)}_0(h_0a)-hJ_0(ha)H'^{(2)}_0(h_0a)\right]}H_0.
\label{b0}\f
If we substitute Eq.~\r{b0} into Eq. (\ref{Ephi}) and use Eq. (\ref{moment}), finally, we derive the magnetic polarizability as
\e
\begin{split}
\alpha^{zz}_{\rm{mm}}=\frac{m_z}{\mu_0H_0}&=\frac{j2k_0^2(\varepsilon_{\rm{r}}-1)}{h_0(k_0^2\varepsilon_{\rm{r}}-k_z^2)}\times\cr
&\frac{\left[ha J_0(ha)-2J_1(ha)\right]}{\left[h_0J_1(ha)H^{(2)}_0(h_0a)-hJ_0(ha)H^{(2)}_1(h_0a)\right]}.
\end{split}
\label{alpha}\f
\begin{figure}[t!]\centering
\subfigure[]{\includegraphics[width=7cm]{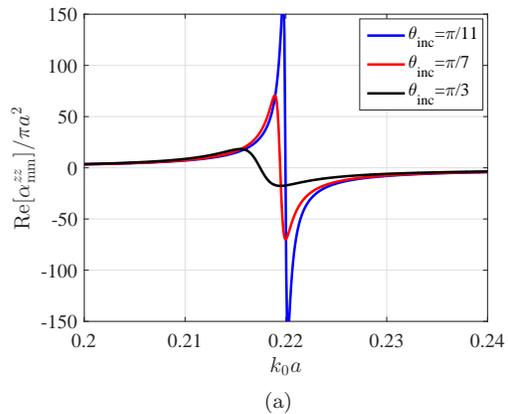}
\label{fig:polarizability1}}
\subfigure[]{\includegraphics[width=7cm]{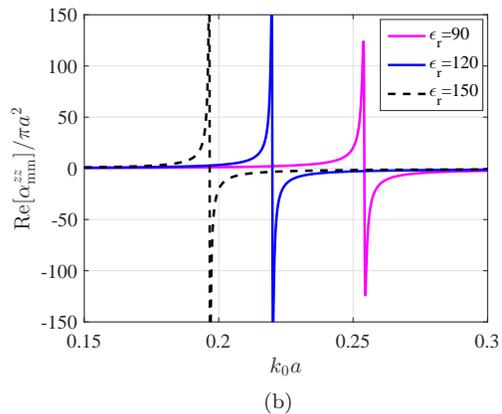}
\label{fig:polar2}}
\caption{(a)--Real part of the magnetic polarizability versus the normalized frequency for three different incident angles. The relative dielectric constant of the dielectric cylinder is assumed to be $\varepsilon_{\rm{r}}=120$. (b)---Real part of the magnetic polarizability versus the normalized frequency for three different dielectric constants. Here, we assume that $\theta_{\rm{inc}}=\pi/11$.}
\end{figure}

To keep only the term with $m=0$ in Eq.~(\ref{Ephi}) is the same as to excite our cylinder by only a magnetic field of the electromagnetic wave letting the external electric field be zero at the cylinder axis. Adding a corresponding term to Eq.~\r{H}, it is easy to see that the excitation of the cylinder by two plane waves incident on it from two opposite sides and forming the standing wave with the node of the electric field at the $z$-axis results in the same formula \r{alpha}. It is worth to note that retaining in Eq.~(\ref{Ephi}) the term $A_1$ we would attribute to the magnetic moment per unit length the dependence on the incident electric field, and the response of the cylinder would be bianisotropic. Keeping only $B_0$ we follow to the approach of \cite{Grimes} that allows one to avoid the seeming bianisotropy in the response of a symmetric scatterer. This bianisotropy is not physically sound and would result in the inconsistencies of the effective-medium model (see e.g. in \cite{Alu}).

Sign minus in the denominator of the second factor in \r{alpha} allows the resonance of the magnetic polarizability if $\varepsilon_{\rm{r}}$ is positive. This is the fundamental Mie resonance of the cylinder.
Since $k_z$ enters both $h$ and $h_0$, the polarizability is non-local i.e. depends on the angle of incidence $\theta_{\rm{inc}}$.
To show the impact of the incidence angle and that of the dielectric constant $\varepsilon_{\rm{r}}$ on the Mie resonance we plot the magnetic polarizability as a function of normalized frequency for several values of $\varepsilon_{\rm{r}}$ and $\theta_{\rm{inc}}$. In Fig.~\ref{fig:polarizability1} it is assumed that $\varepsilon_{\rm{r}}=120$. Then the Mie resonance occurs approximately at $k_0a=0.22$  for all angles. For small angles, the resonance is, indeed, stronger than that for large angles. Thereby, we see that the resonant polarizability depends strongly on the incident angle or -- more specifically -- on the axial component $k_z=k_0\cos(\theta_{\rm{inc}})$ of the wave vector.

At frequencies below the Mie resonance band the magnetic polarizability is negligibly small. The location of the resonance on the frequency axis is a function of the dielectric constant of the cylinder, as can be seen in Fig.~\ref{fig:polar2}. When $\varepsilon_{\rm{r}}$ is not very high, the resonance happens at higher $k_0a$ and vice versa. For the accuracy of the effective-medium model it is better if the resonance occurs at lower $k_0a$.
This model requires that both the wires radius ($a$) and the array period ($b$) are sufficiently small compared to the wavelength in host medium, in other words, $k_0a<k_0b\ll 1$. Materials such as lithium tantalate, silicon carbide, or hexagonal boron nitride which support phonon polaritons would provide such high dielectric constants at the infrared. The only problem with these materials may be optical losses which are obviously noticeable when $\varepsilon_{\rm{r}}\gg 1$. 
\begin{figure}[t!]\centering
\subfigure[]{\includegraphics[width=7cm]{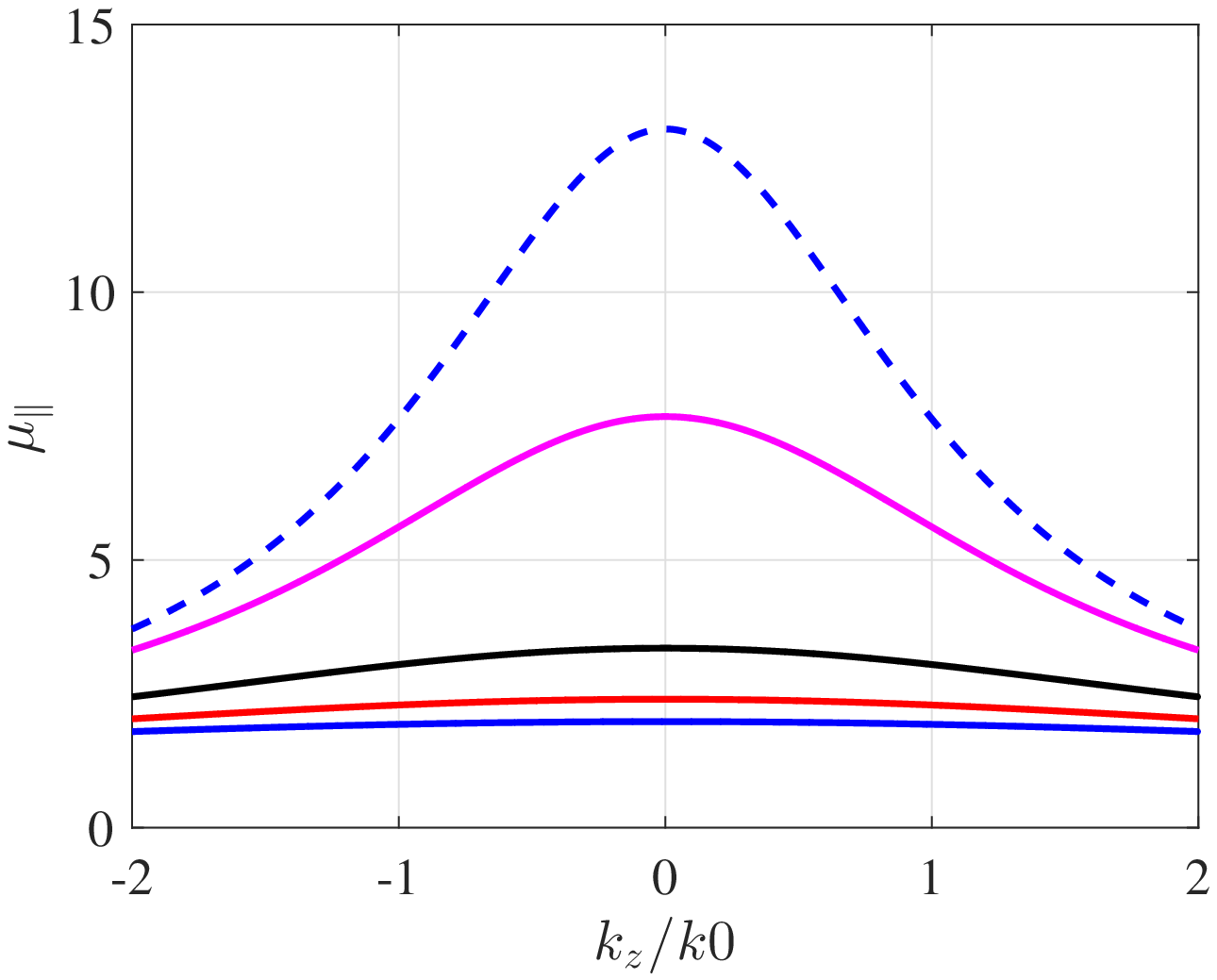}}
\subfigure[]{\includegraphics[width=7cm]{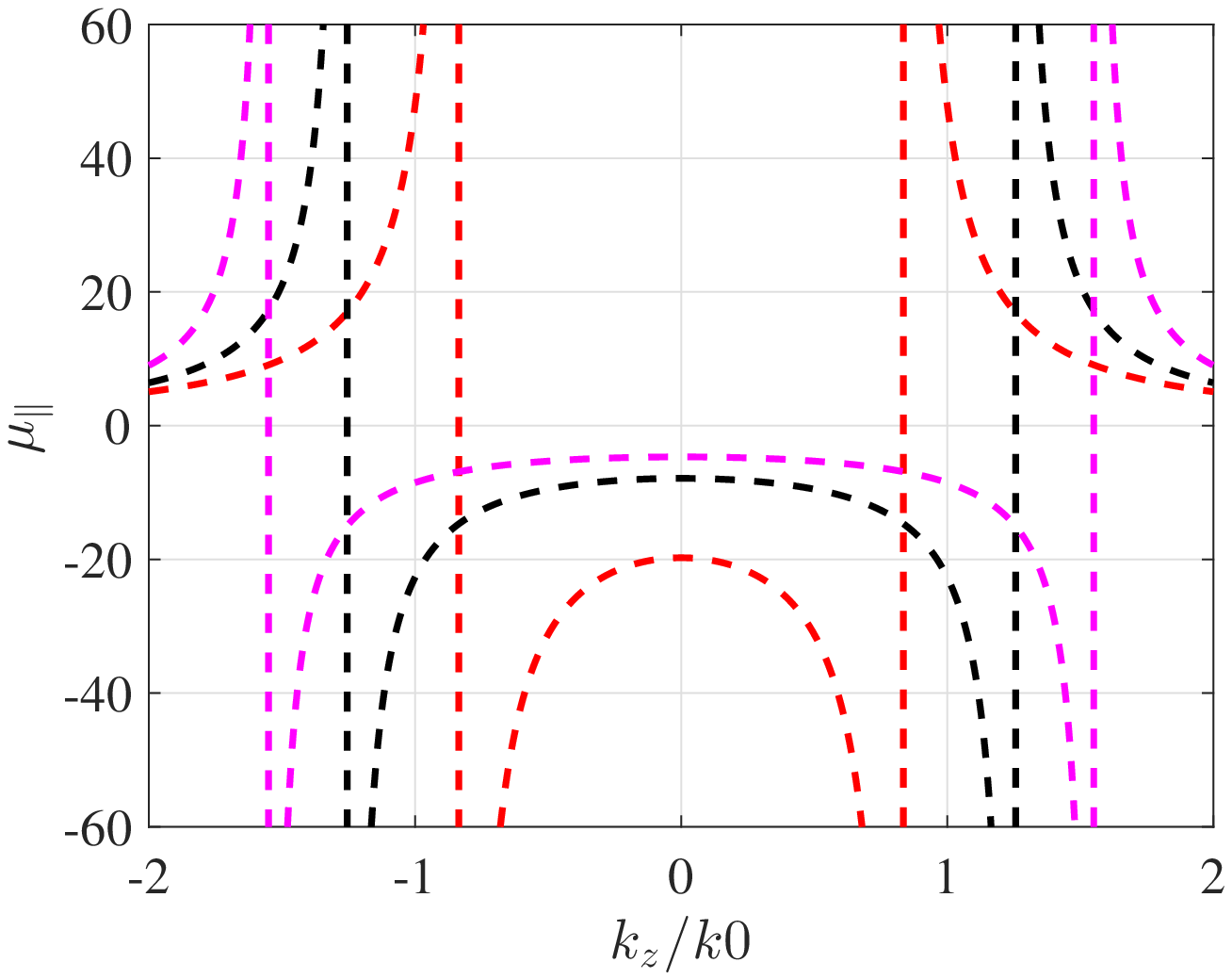}}
\caption{The parallel component of the effective permeability versus the normalized parallel component of the wave vector, when
$\varepsilon_{\rm{r}}=120$ and $f_{\rm{v}}=0.1963$.
(a)--Frequencies below the topological transition $k_0a=0.22$.
Solid blue curve--$k_0a=0.2058$, solid red one--$k_0a=0.2096$, solid black one--$k_0a=0.2135$, solid magenta one--$k_0a=0.2173$, and dashed blue one--$k_0a=0.2182$. (b)--
Frequencies above the topological transition $k_0a=0.22$. Dashed red--$k_0a=0.2201$, dashed black--$k_0a=0.2211$, dashed magenta--$k_0a=0.2220$. }
\label{fig:perm}
\end{figure}

Deriving the effective permeability of the lattice we follow to the approach \cite{mario2,mario1} which allows us to take into account the non-locality also in the electromagnetic interaction of cylinders.
For the axial component of the effective permeability we have the following formula:
\begin{equation}
\mu_\parallel=1+\frac{1}{A_{\rm{cell}}}(\frac{1}{\alpha^{\rm{zz}}_{\rm{mm}}}-C_{\rm{int}}^z)^{-1},
\end{equation}
 which is dual to that obtained for the axial permittivity in \cite{mario2} with the substitution of the magnetic polarizability by the electric one.
Here $A_{\rm{cell}}=b^2$ and $C_{\rm{int}}^z$ is the lattice interaction constant responsible for the dipole-dipole interaction. It is the same for magnetic and electric dipole moment per unit length of the cylinder and
 was derived in \cite{mario2}:
\begin{equation}
\begin{split}
C_{\rm{int}}^z\approx &h_0^2\displaystyle\left(\frac{j}{4}+\frac{1}{2\pi}\ln(\frac{h_0b}{4\pi})+\xi\right),\cr
&\xi=\frac{C}{2\pi}+\frac{1}{12}+\sum_{n=1}^\infty\frac{(e^{2\pi\vert n\vert-1})^{-1}}{\pi\vert n\vert}.
\end{split}
\end{equation}
Here, $C$ is the Euler constant. Let us choose $\varepsilon_{\rm{r}}=120$ and the fraction volume $f_{\rm{v}}=\pi a^2/b^2=0.1963$ (this corresponds to the relation between the period and the wire radius $b=4a$). In Fig.~\ref{fig:perm} we depict the axial component of the effective permeability  as a function of $k_z$ for these design parameters. For very low $k_0a\ll0.22$, $\mu_\parallel(k_z)$ is approximately uniform and close to unity. The increase of $k_0a$ results in increasing $\mu_\parallel(k_z)$ especially for the transverse propagation, however it keeps finite and smooth versus $k_z$. At frequency $k_0a=0.22$, $\mu_\parallel (k_z)$ changes its behavior
that corresponds to the topological transition. For $k_0a=0.2201$ the axial permeability is negative in the interval $\vert k_z/k_0\vert<0.83$ and becomes positive in $\vert k_z/k_0\vert>0.83$. At $\vert k_z/k_0\vert=0.83$, there is a vertical asymptote. This resonance of $\mu_\parallel (k_z)$ implies that the transverse component $q$ of the wave vector experiences the similar resonance at $\vert k_z/k_0\vert=0.83$. It is clear from the
dispersion equation for TE-waves in a magnetic hyperbolic medium \cite{mario1} (for the case when the host medium is free space $\varepsilon_{\rm{h}}=1$):
\begin{equation}
{q^2/\mu_\parallel (k_z)}+{k_z^2/\mu_{\perp}}=k_0^2, \quad \mu_{\perp}\approx 1.
\end{equation}
In practical cases -- when optical losses are taken into account -- the resonance of $q$ means that $q$ is a very large imaginary value. As we can see in Fig.~\ref{fig:perm} this effect keeps for propagating eigenmodes within a certain range of frequencies. For frequencies $k_0a\ge 0.23$ the resonances of $\mu_\parallel (k_z)$ still hold but correspond to very high spatial frequencies. For so high spatial frequencies the effective-medium model is not applicable \cite{Kidwai}. In our plot Fig.~\ref{fig:perm} the region of $k_z$ (of the order of $k_0$ or smaller) is that compatible with the model.

Imaginary $q$ for a propagating eigenmode with allowed value of $k_z$ means that the medium eigenmode propagates along the $z$-axis, being evanescent in the transverse plane $(x-y)$. Very high imaginary $q$
implies the ultimate subwavelength concentration of the eigenmode in this plane. If we embed a subwavelength source of TE-waves (such as a point magnetic dipole oriented orthogonally to the $z$-axis) in our wire medium this mode will be excited dominating over all other modes propagating along the $z$-axis. This domination results from the huge slope of the curve $\mu_\parallel (k_z)$ which implies the huge density of electromagnetic states. This feature was noticed for the axial permittivity of some dielectric hyperbolic metamaterials in \cite{Shalaev}. Our magnetic hyperbolic metamaterial is dual to dielectric hyperbolic metamaterials and the roles of $\overline{\overline{\epsilon}}$ and TM-waves are played by $\overline{\overline{\mu}}$ and TE-waves. Since the dominant mode strongly attenuates in both $x$- and $y$-directions, it means that the whole electromagnetic field generated by our magnetic dipole
should be concentrated around the line which passes through the dipole center along the $z$-axis.

\section{Numerical simulations}
\label{sec:pf}
\subsection{Localization of magneto-dipole radiation}

\begin{figure}[t!]\centering
{\includegraphics[width=3.75cm]{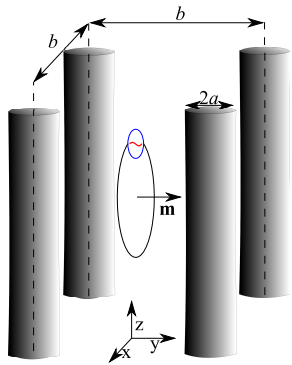}}
\caption{The subwavelength magnetic dipole oriented perpendicularly to the wire axis and located in the center of the wire medium sample.}
\label{fig:mdm}
\end{figure}
\begin{figure}[t!]\centering
\subfigure[$f=18$ THz ($k_0a=0.1715$)]{\includegraphics[width=4.16cm]{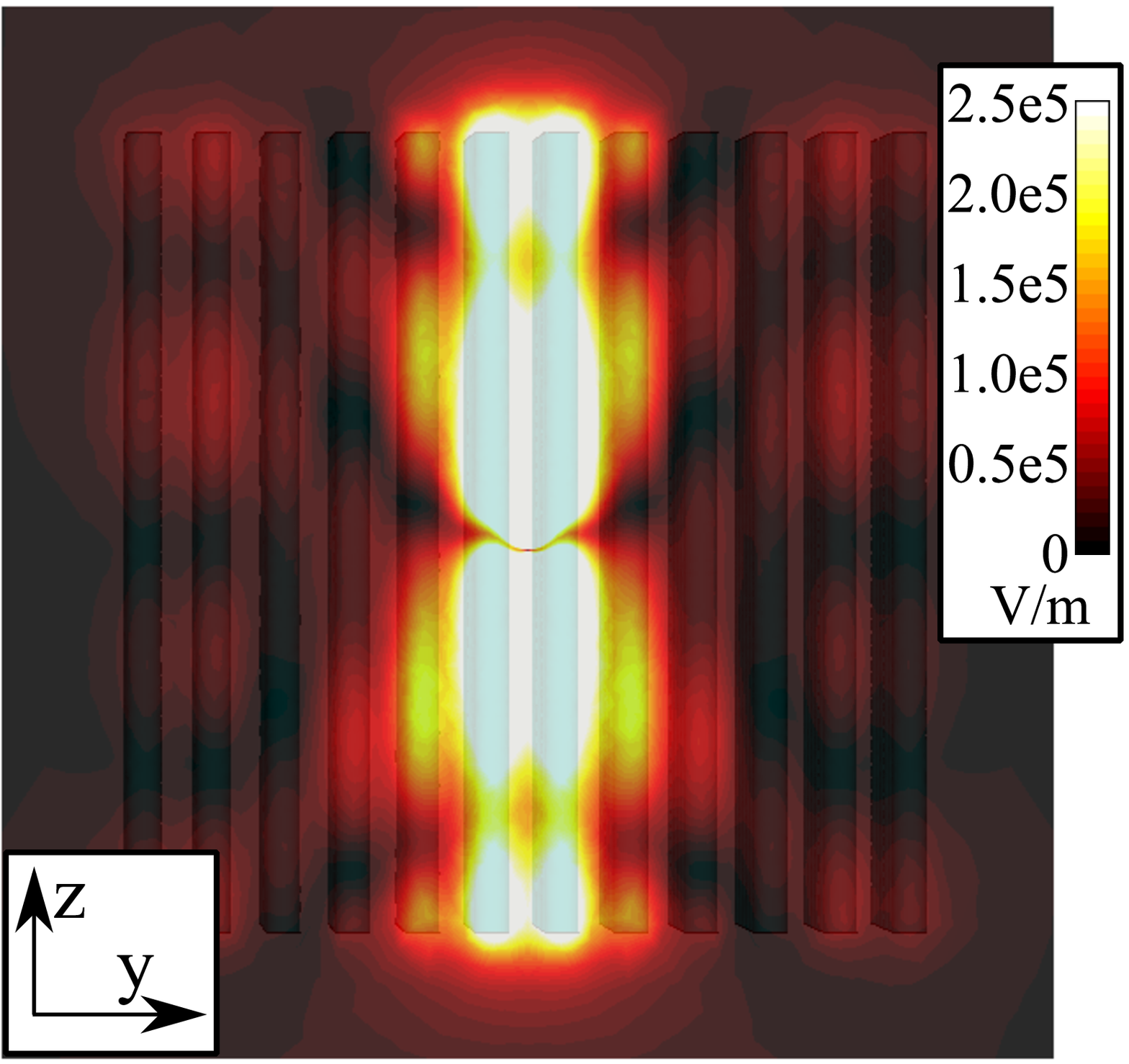}
\label{fig:FD1}}
\subfigure[$f=21$ THz ($k_0a=0.2001$)]{\includegraphics[width=4.16cm]{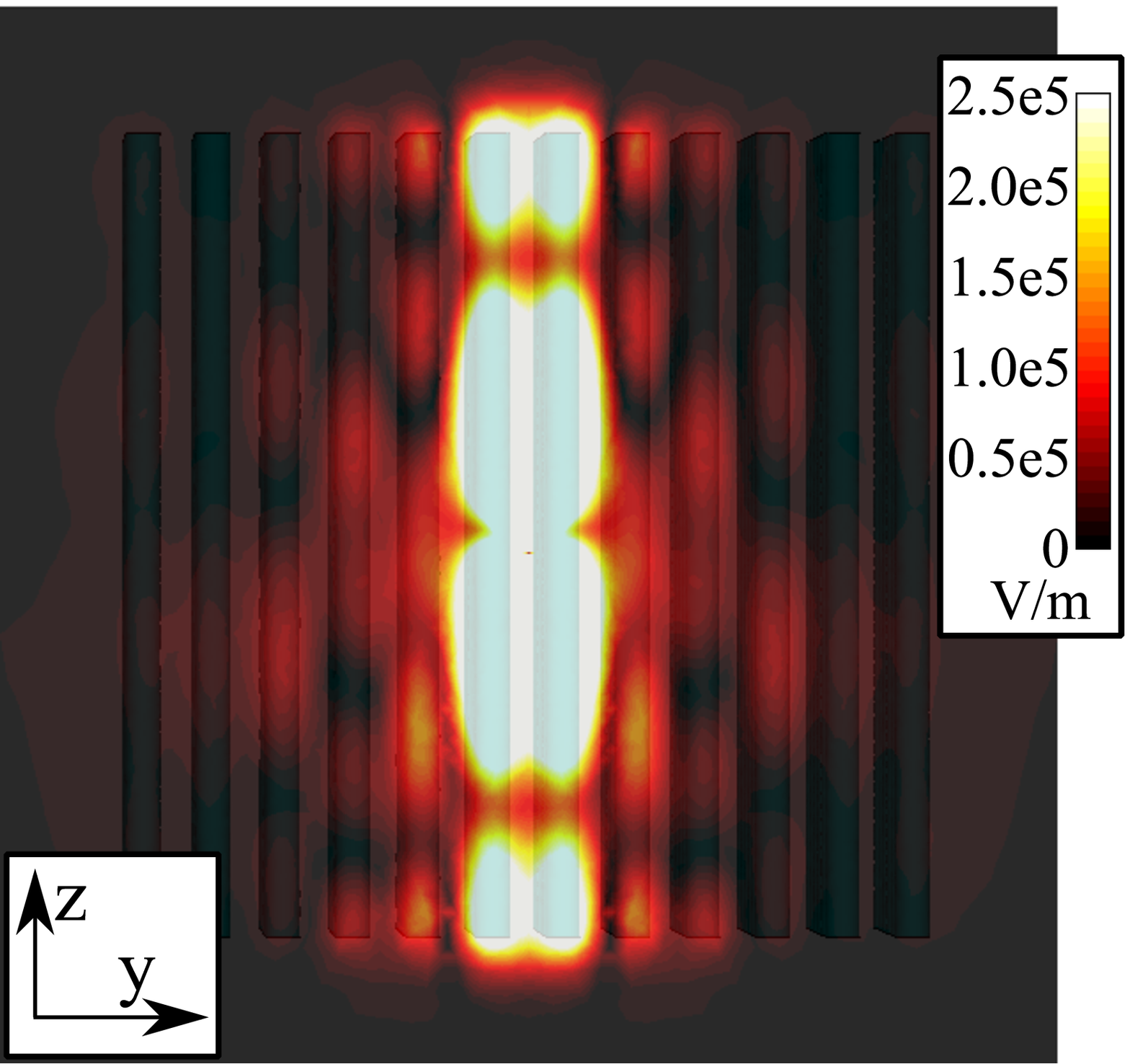}
\label{fig:FD2}}
\subfigure[$f=23$ THz ($k_0a=0.2192$)]{\includegraphics[width=4.16cm]{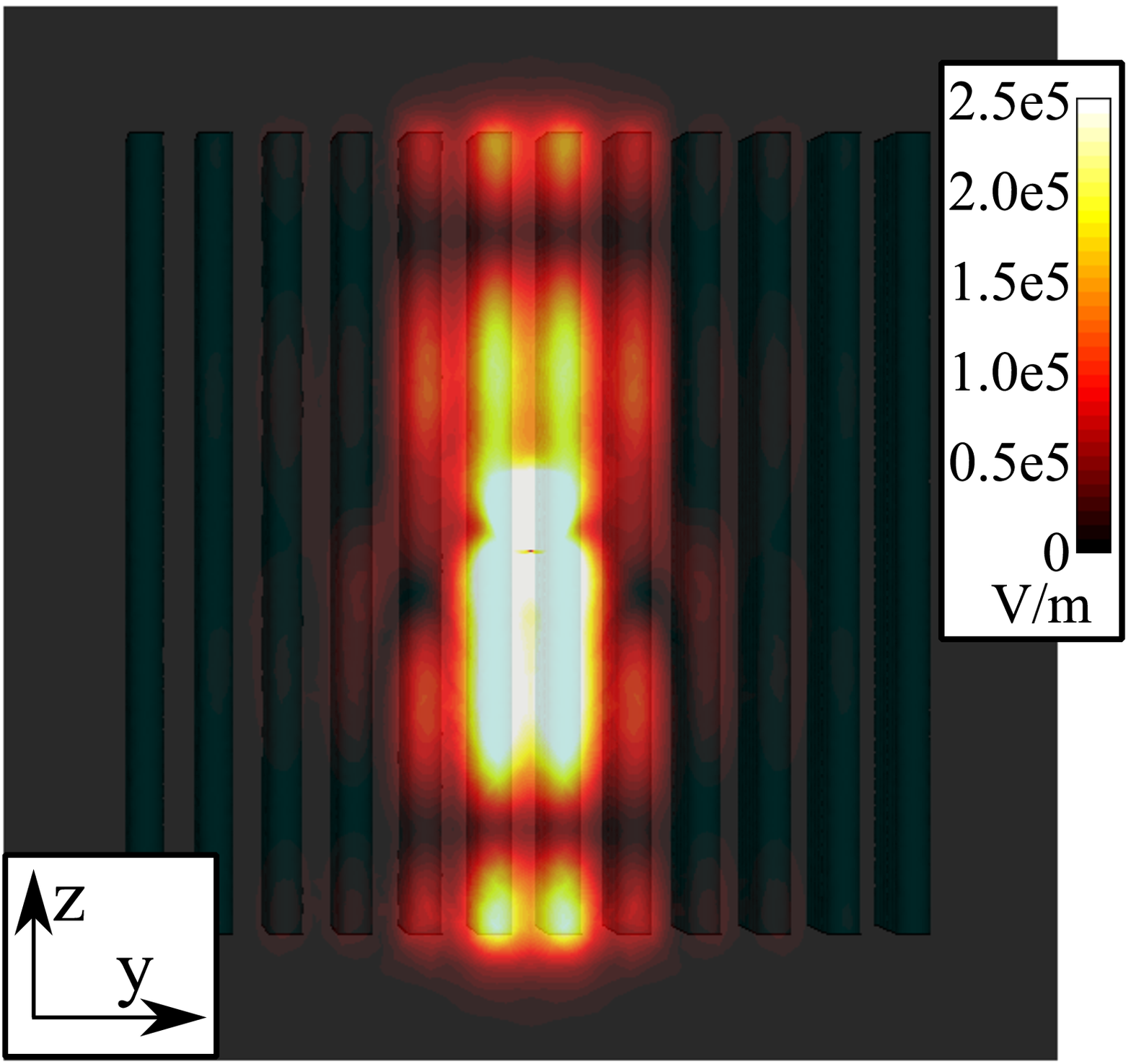}
\label{fig:FD3}}
\subfigure[$f=27$ THz ($k_0a=0.2573$)]{\includegraphics[width=4.16cm]{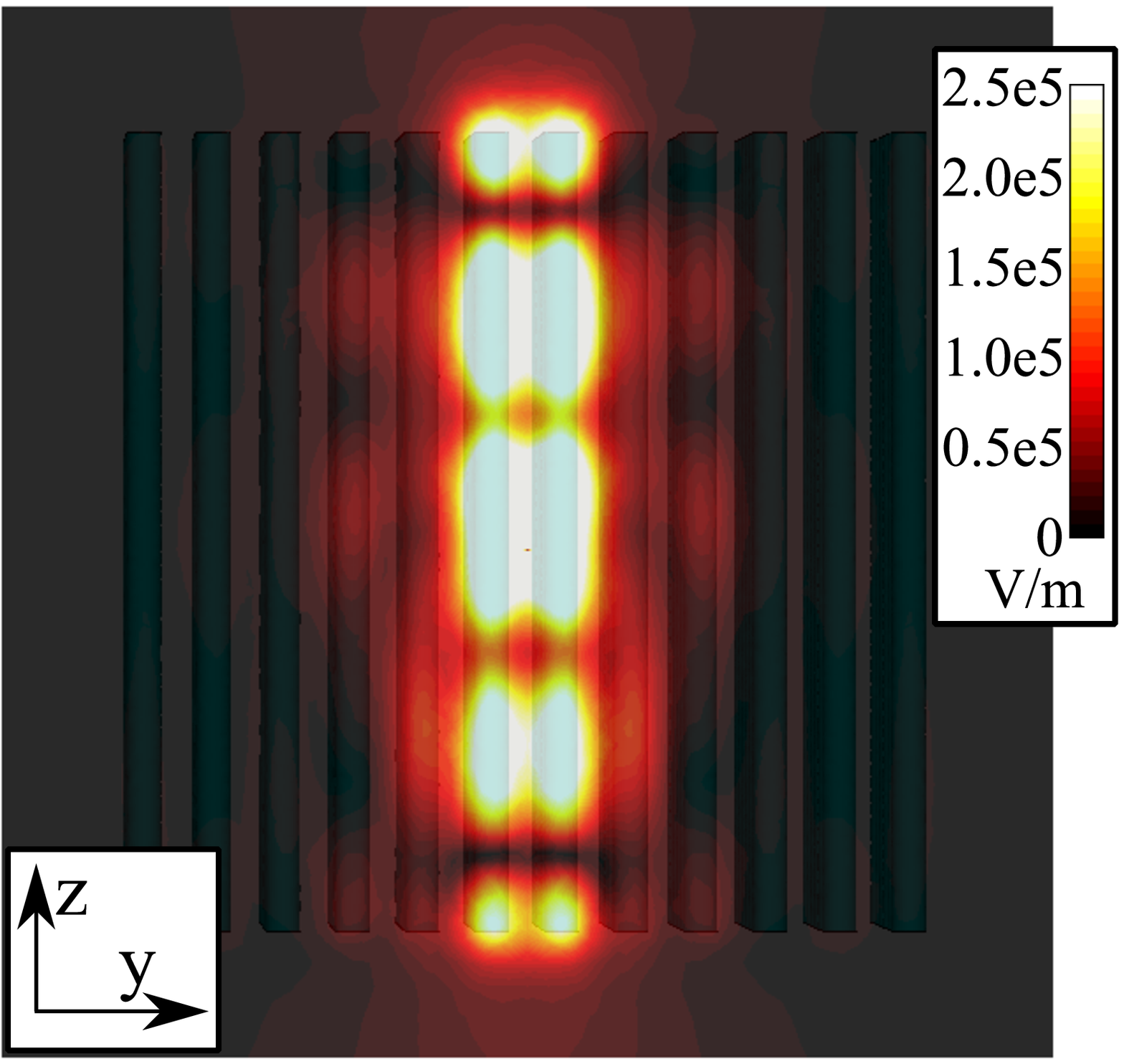}
\label{fig:FD4}}
\caption{Distribution of the normal component of the electric-field in the wire-medium sample (consisting of $12\times12$ nanowires). The magnetic dipole moment located in the center of the sample is along the $\mathbf{y}_0$-axis. Here, $\varepsilon_{\rm{r}}=120-j12$, $f_{\rm{v}}=0.1963$.}
\label{fig:efTE}
\end{figure}

This filamentary localization of magneto-dipole radiation was confirmed by using a 3D electromagnetic simulator CST Microwave Studio. In our simulations we position a strongly subwavelength current loop whose magnetic moment is directed along the $y$-axis in the center of the wire medium sample as shown in Fig.~\ref{fig:mdm}.  The medium sample is finite-size with length $L_{\rm{sample}}=(N-1)b+2a$ where $N=12$ is the number of the wires.
It has realistic optical losses: we suppose that the tangent of dielectric losses in the material of our wires is equal to $\tan\delta=0.1$, where the complex permittivity of nanowires is $\varepsilon_{\rm{r}}={\rm{Re}}[\varepsilon_{\rm{r}}][1-j\tan\delta]$, ${\rm{Re}}[\varepsilon_{\rm{r}}]=120$. This value is the complex permittivity of lithium tantalate at $f=23$ THz.
The radius of the wires is $a=455$ nm and $b=4a$ as above.  The spatial distribution of the transverse component of the electric field (the $x$-component in the $y-z$ plane) is shown in Fig.~\ref{fig:efTE} for four frequencies. We see that only at the frequency $k_0a=0.2192\approx0.22$, the field so significantly decays in the transverse plane so that it is practically concentrated in one unit cell of the wire medium. This is the maximal possible concentration in a composite material -- that restricted only by the granularity of the medium. The excited wave propagates along the $z$-axis and experiences nearly total internal reflection at the interfaces of the medium sample. Similar pictures correspond to other components of the electromagnetic field. The comparison of different components confirms that the wave is TE-polarized. At three other frequencies the waves are also TE-polarized, whereas their localization is not filamentary and even the cross-like dipole pattern inherent to usual hyperbolic metamaterials \cite{poddubny2} can be guessed in Fig.~\ref{fig:FD1}.

The reason of the striking difference in the internal dipole patterns for our wire medium and the majority of hyperbolic materials is spatial dispersion.
Metal wire media operating at microwaves are also spatially dispersive, and the slope of $\epsilon_\parallel(k_z)$ (see e.g.
in \cite{simovski1}) at their spatial resonance $|k_z|=k_0$ is as huge as that of our $\mu_\parallel(k_z)$ at their resonant $k_z$ in the topological transition frequency range.
It is not surprising, therefore, that in microwave wire media the similar filamentary dipole pattern was also observed (see e.g. in \cite{nefedov}).
However, in what concerns this pattern our infrared magnetic wire medium and microwave wire media differ qualitatively.
The first difference is duality: microwave wire media are hyperbolic dielectric metamaterials and the filamentary pattern in them corresponds to transverse \emph{electric} dipoles.
The second difference is the effect bandwidth: microwave wire media are broadband, and the same dipole pattern is inherent for them at any frequency for which the effective-medium model is applicable.
Our wire media possess the dipole pattern which strongly varies versus frequency, that is illustrated by Fig.~\ref{fig:efTE}. The filamentary pattern of the magnetic dipole corresponds only to the
range of the topological transition.

\subsection{Purcell factor}

\begin{figure}[t!]\centering
\subfigure[]{\includegraphics[width=7cm]{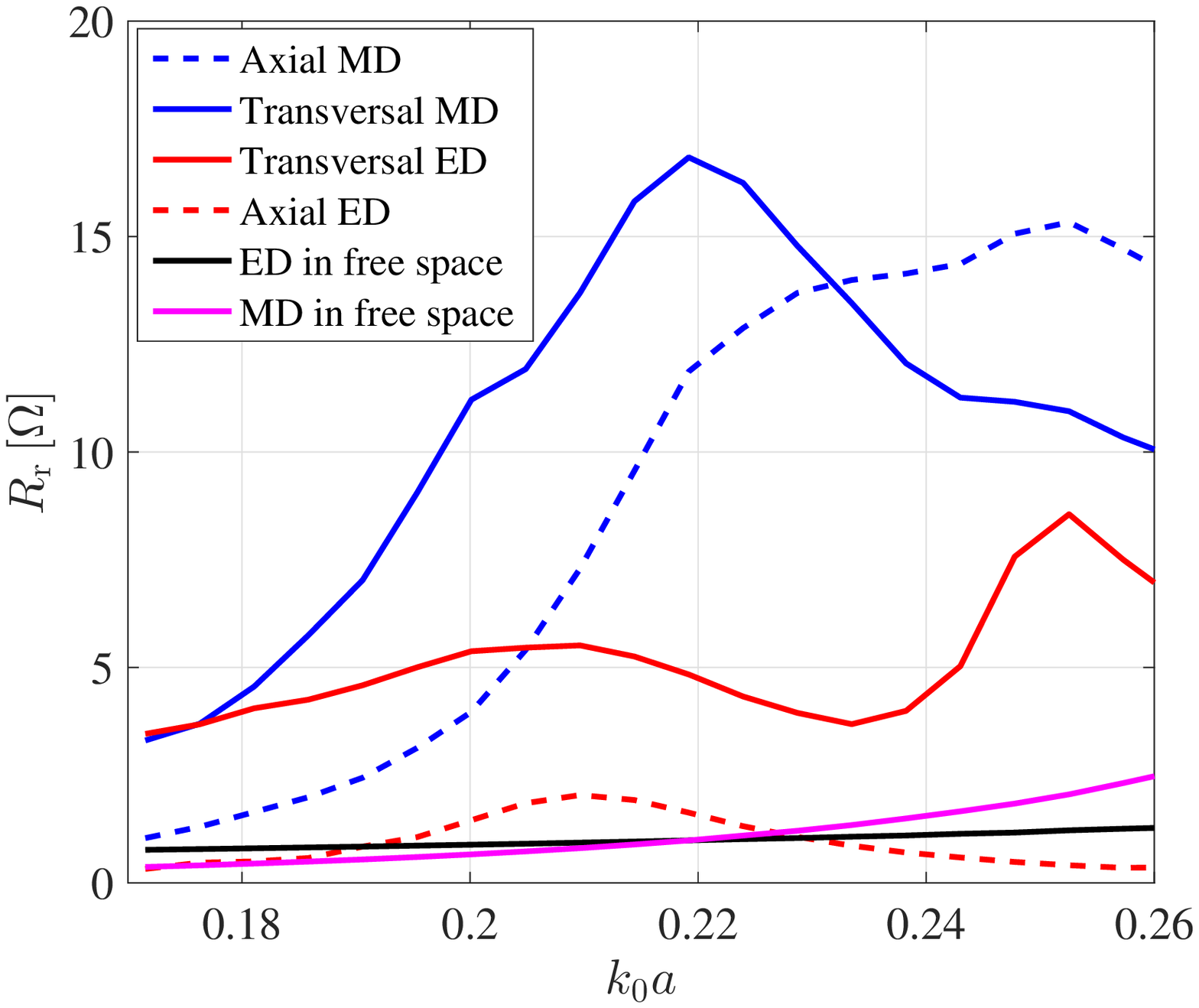}
\label{fig:RE1}}
\subfigure[]{\includegraphics[width=6.9cm]{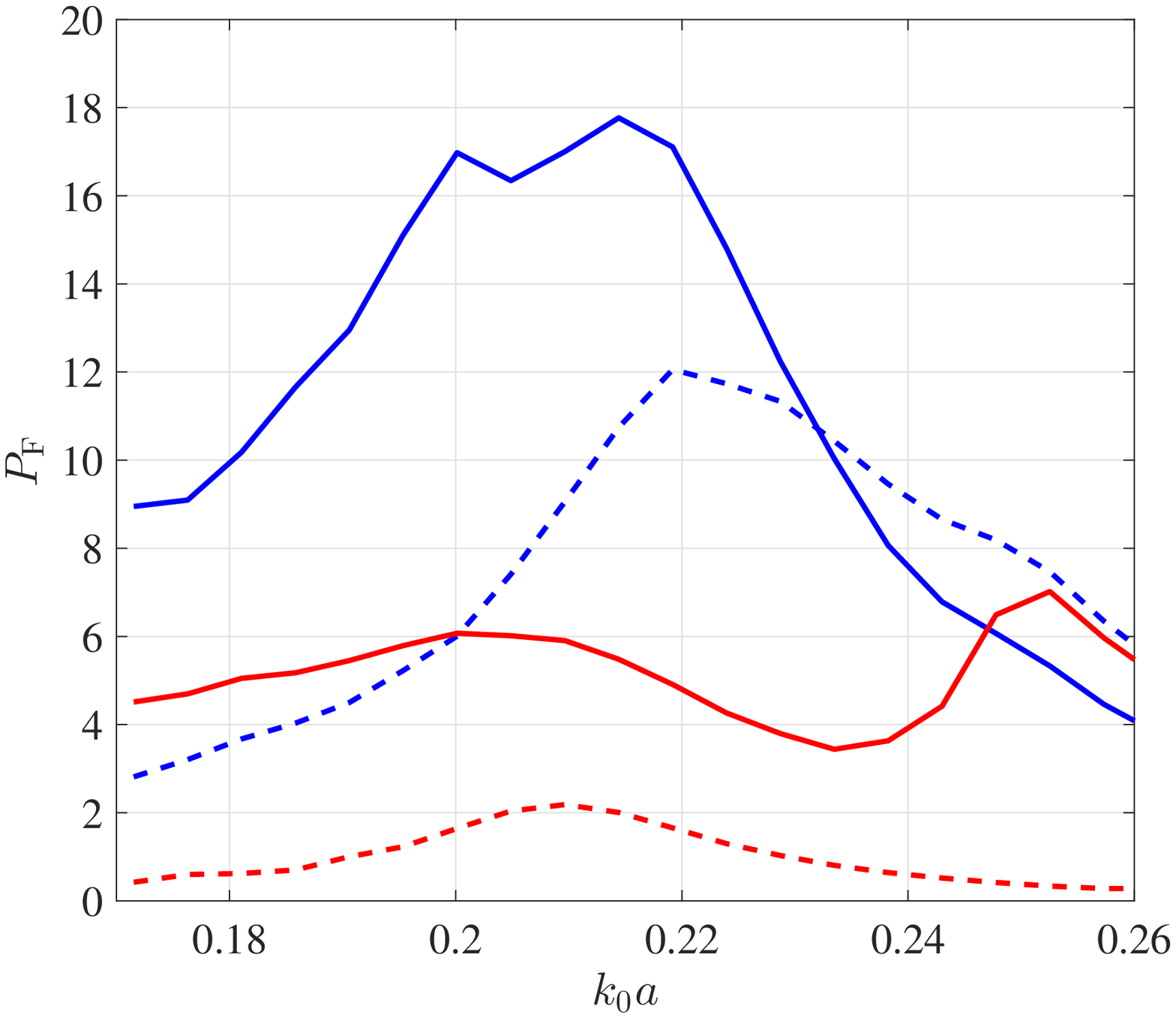}
\label{fig:RE2}}
\caption{(a)--Radiation resistance of subwavelength magnetic (MD) and electric (ED) dipoles located in the center of the wire-medium sample (including $12\times12$ wires) and the free space. (b)--Corresponding Purcell factor. Here, $\varepsilon_{\rm{r}}=120-j12$ and $f_{\rm{v}}=0.1963$.}
\end{figure}

Another implication of the topological transition in our magnetic metamaterial is the strong Purcell effect. Originally, the effect was known as the enhancement of the decay rate of a quantum emitter
located in an open cavity \cite{purcell,novotny}. However, the role of the cavity is only to extract more power from the emitter, whereas
the enhancement of its decay rate is the same as the enhancement of the radiated power. Therefore, the notion of the Purcell effect was extended, first, to
any scatterer located in the vicinity of an emitter \cite{sauvan,pelton,tam,anger}, and, finally to any active radiator whose radiation is enhanced by any environment different from free space \cite{jacob,krasnok,mirmoosa2}.

Based on this general concept, we can define the magnetic Purcell factor of our wire medium as the increase of the radiated power of a subwavelength magnetic dipole due to the presence of the wire medium around it.
According to the \textquotedblleft antenna terminology\textquotedblright, the radiated power of the emitter is proportional to the radiation resistance of that emitter \cite{balanis}. Therefore, to calculate the Purcell factor, it is the same as to obtain the ratio of the radiation resistances in presence of the wire medium ($R_{\rm{r}}$) and in absence ($R_{{\rm{r}}_0}$):
\begin{equation}
P_{\rm{F}}=\displaystyle\frac{R_{\rm{r}}}{R_{{\rm{r}}_0}}.
\end{equation}
For the same structure as above we have simulated the radiation resistances. Besides of the case of the transverse magnetic dipole shown in Fig.~\ref{fig:mdm}, we have studied the case when the same magnetic dipole is oriented in parallel to the $z$-axis. Also, we simulated the cases of the transverse and axial electric dipoles of the same subwavelength size, all centered at the same point as in Fig.~\ref{fig:mdm}.
In Fig.~\ref{fig:RE1} which depicts the radiation resistance versus frequency, the blue curves refer to the magnetic dipole in presence of the sample. As one can see, the radiation resistance of the magnetic dipole oriented perpendicularly to the optical axis experiences the strong resonance at $k_0a\approx0.22$, exactly where the effective-medium model predicts the topological transition. The dashed blue curve shows weaker effect for the parallel magnetic dipole. Red solid and dashed curves correspond to the perpendicular and parallel electric dipoles, respectively. For comparison with the case when the host is free space, the frequency dispersion of $R_{{\rm{r}}_0}$ is given in the same plot for both electric and magnetic dipoles. Figure~\ref{fig:RE2} presents the frequency dispersion of the Purcell factor for all four dipoles.
It is clearly seen that the strongest resonance and highest $P_{\rm{F}}$ correspond to the transversal magnetic dipole. An axial magnetic dipole also creates the TE-polarized spatial spectrum, and therefore it also
experiences the resonance at the topological transition. However, the resonant values of $P_{\rm{F}}$ are smaller because it mainly creates the TE-polarized radiation which weakly interacts with the nanowires i.e. is not enhanced.

The values of $P_{\rm{F}}$ for the axial electric dipole are low. It is not surprising. Recall, that in a wire medium of perfectly conducting (PC) wires the radiation of the axial electric dipole is fully suppressed \cite{poddubny1}.
The suppression results from the destructive near-field coupling of the dipole with adjacent wires. The axial current of the dipole induces opposite axial currents in them which cancel the radiation.
This effect is similar to the suppression of radiation of a horizontal electric dipole located on the PC substrate. Well, our wires are not PC, they are highly refractive. However, the substrates
with high positive permittivity and finite negative permittivity also suppress the radiation of the horizontal electric dipole. This destructive interaction origins from the capacitive coupling
which causes the inverse mirror image of the horizontal electric dipole in the substrate with the strong skin-effect. A similar situation holds for the axial dipole source in wire media if the wires possess strong skin-effect.
If the wires have high or negative $\varepsilon_{\rm r}$, the values of $P_{\rm{F}}$ for the axial electric dipole should be low. However, we can see in Figure~\ref{fig:RE2} that the radiation of this dipole experiences the resonance (at slightly lower frequency than that of the magnetic topological transition). This is also not surprising, because not all radiation of an axial electric dipole is TM-polarized. It also produces some TE-waves,
and this effect is competing with the capacitive destructive interaction. Therefore, we have $P_{\rm{F}}=2$ at $k_0a=0.21$, whereas far from this frequency we see $P_{\rm{F}}\approx 0$.
As to the noticeable enhancement of the transverse electric dipole, it, definitely, occurs due to the constructive interaction of this dipole with the adjacent nanowires (similarly to a vertical dipole on the high-permittivity substrate). This effect is complemented by the resonant enhancement of TE-polarized waves. However, in the near zone of the electric dipole the magnetic field is weak. Therefore, an electric dipole cannot interact with the
magnetic medium as strongly as a magnetic dipole, and the resonant Purcell factor of the transversal magnetic dipole is higher.

\section{Conclusions}
\label{sec:con}
In this work we calculated the effective permeability of a simple wire medium whose wires are made of a material with high positive dielectric constant.
Its parallel component $\mu_\parallel$  is resonant due to the Mie resonance of a single wire, and is an indefinite function of both frequency and axial wave vector $k_z$
(changes the sign depending on these arguments). This change of the sign results in a transition in the topology of the dispersion surface for TE polarized waves.
In the frequency range of the topological transition, the huge (filamentary) localization of radiation occurs for an internal magnetic source, that clearly shares our medium out from other hyperbolic metamaterials, including
that introduced in \cite{mario1}. Though a similar pattern is inherent to the medium of perfectly conducting wires operating at microwaves, in our case this pattern corresponds to the magnetic dipole and occurs only
at the topological transition.

Our wire medium manifests a strong radiation enhancement for an internal dipole source. Unlike dielectric hyperbolic metamaterials, including microwave wire media,
and polaritonic media operating at higher frequencies \cite{mirmoosa2} the maximal enhancement corresponds to magnetic dipoles. Unlike spiral media \cite{mario1} it holds at the topological transition
and the maximal radiation enhancement corresponds to the transversal magnetic dipole. 

What is also important: from the comparison of the present paper with \cite{mirmoosa2} it is clear that the same polaritonic wire media which manifest the resonant effects related to the regime \emph{epsilon-near-zero} at 35--45 THz \cite{mirmoosa2} may manifest the similar (but different) resonant effects related to the regime \emph{mu-near-zero} in the range 15--30 THz. This is another argument in favor of this type of hyperbolic metamaterials, which definitely deserve more attention of theorists and an experimental implementation to confirm the claimed effects.


\begin{thebibliography}{99}


\bibitem{kabashin}
A.~V. Kabashin, P.~Evans, S.~Pastkovsky, W.~Hendren, G.~A.~Wurtz, R.~Atkinson, R.~Pollard,
V.~A.~Podolskiy, and A.~V.~Zayats,
Plasmonic nanorod metamaterials for biosensing,
Nat.~Mater.~\textbf{8}, 867--872 (2010).

\bibitem{belov1}
Y. Zhao, G. Palikaras, P.~A. Belov, R.~F. Dubrovka, C.~R Simovski, Y. Hao, and C.~G. Parini,
Magnification of subwavelength field distributions using a tapered array of metallic wires with planar interfaces and an embedded dielectric phase compensator,
New J. Phys. \textbf{12}, 103045 (2010).

\bibitem{casse}
B.~D.~F.~Casse, W.~T.~Lu, Y.~J.~Huang, E.~Gultepe, L.~Menon, S.~Sridhar,
Super-resolution imaging using a three-dimensional metamaterials nanolens,
Appl.~Phys.~Lett.~\textbf{96}, 023114 (2010).

\bibitem{wan}
Z.-K. Zhou, M. Li, Z.-J. Yang, X.-N. Peng, X.-R. Su, Z.-S. Zhang, J.-B. Li, N.-C. Kim,
X.-F. Yu, L. Zhou, Z.-H. Hao, and Q.-Q. Wan, Plasmon-mediated radiative energy transfer across a silver nanowire array
via resonant transmission and subwavelength imaging, ACS Nano \textbf{4} 5003–-5010 (2010).


\bibitem{poddubny1}
A.~N.~Poddubny, P.~A.~Belov and Y.~S.~Kivshar,
Purcell effect in wire metamaterials,
Phys.~Rev.~B~\textbf{87}, 035136 (2013).

\bibitem{mirmoosa2}
M.~S.~Mirmoosa, S.~Yu.~Kosulnikov and C.~R.~Simovski,
Unbounded spatial spectrum of propagating waves in a polaritonic wire medium,
Phys.~Rev.~B~\textbf{92}, 075139 (2015).

\bibitem{simovski}
C. Simovski, S. Maslovski, I. Nefedov, S. Kosulnikov, P. Belov, and S. Tretyakov,
Hyperlens makes thermal emission strongly super-Planckian,
Photonics~and~Nanostructures: Fundamentals~and~Applications \textbf{13}, 31--41 (2015).


\bibitem{simovski2}
I.~S.~Nefedov and C.~R.~Simovski,
Giant radiation heat transfer through micron gaps,
Phys.~Rev.~B~\textbf{84}, 195459 (2011).



\bibitem{mirmoosa1}
M.~S.~Mirmoosa and C.~R.~Simovski,
Micron-gap thermophotovoltaic systems enhanced by nanowires,
Photon.~Nanostruct.~Fundam.~Appl.~\textbf{13}, 20--30 (2015).



\bibitem{simovski1}
C.~R.~Simovski, P.~A.~Belov, A.~V.~Atrashchenko and Y.~S.~Kivshar,
Wire metamaterials: physics and applications,
Adv.~Mater.~\textbf{24}, 4229--4248 (2012).

\bibitem{mario2}
M.~G.~Silveirinha,
Nonlocal homogenization model for a periodic array of $\epsilon$-negative rods,
Phys.~Rev.~E~\textbf{73}, 046612 (2006).

\bibitem{poddubny2}
A.~Poddubny, I.~ Irosh, P.~Belov and Y.~Kivshar,
Hyperbolic metamaterials,
Nature~Photonics~\textbf{7}, 948--957 (2013).

\bibitem{mario1}
T.~A.~Morgado, J.~T.~Costa and M.~G.~Silveirinha,
Magnetic uniaxial wire medium,
Phys.~Rev.~B~\textbf{93}, 075102 (2016).

\bibitem{spiral}
C.~R. Simovski, Dispersion properties of metal photonic crystals form parallel spirals, in: \emph{Advances in Electromagnetics of Complex Media and Metamaterials}, S. Zouhdi, A. Sihvola, and M. Arsalane (editors),
Kluwer Academic Publishers (2002), pp. 207-227.

\bibitem{spiral1}
P.~A. Belov, C.~R. Simovski, and S.~A. Tretyakov, An example of bi-anisotropic electro-magnetic crystals: the spiral medium, Phys. Rev. E \textbf{67}, 056622 (2003).

\bibitem{Grimes}
C.A. Grimes and D.M. Grimes, Permeability and permittivity spectra of granular materials, Phys. Rev. B \textbf{43}, 10780-10788 (1991).

\bibitem{Alu}
A. Al{\`u},
Restoring the physical meaning of metamaterial constitutive parameters,
\textit{Phys. Rev. B} \textbf{83}, 081102(R) (2011).

\bibitem{Kidwai}
O. Kidwai, S.~V. Zhukovsky and J.~E. Sipe,
Dipole radiation near hyperbolic metamaterials: applicability of effective-medium approximation, Opt. Lett. \textbf{36}, 2530–-2532 (2011).

\bibitem{Shalaev}
Z.~Jacob, J.~Y.~Kim, G.~V.~Naik, A.~Boltasseva, E.~E.~Narimanov and V.~M.~Shalaev, 
Engineering photonic density of states using metamaterials, 
Appl.~Phys.~B~\textbf{100}, 215--220 (2010).

\bibitem{nefedov}
I.~S.~Nefedov and A.~J.~Viitanen, Effective-medium approach for subwavelength resolution, Electron.~Lett.~\textbf{43}, 22--25 (2007).

\bibitem{purcell}
E.~M.~Purcell,
Spontaneous emission probabilities at radio frequencies,
Phys.~Rev.~\textbf{69}, 681 (1946).

\bibitem{novotny}
L.~Novotny and B.~Hecht, \emph{Principles of Nano-Optics} (Cambridge University Press, Cambridge, UK, 2006).

\bibitem{sauvan}
C.~Sauvan, J.~P.~Hugonin, I.~S.~Maksymov and P.~Lalanne,
Theory of the spontaneous optical emission of nanosize photonic and plasmon resonators,
Phys.~Rev.~Lett.~\textbf{110}, 237401 (2013).

\bibitem{pelton}
M.~Pelton,
Modified spontaneous emission in nanophotonic structures,
Nature~Photonics~\textbf{9}, 427--435 (2015).

\bibitem{tam}
F.~Tam, G.~P.~Goodrich, B.~R.~Johnson and N.~J.~Halas,
Plasmonic enhancement of molecular fluorescence,
Nano~Lett.~\textbf{7}, 496--501 (2007).

\bibitem{anger}
P.~Anger, P.~Bharadwaj and L.~Novotny,
Enhancement and quenching of single-molecule fluorescence,
Phys.~Rev.~Lett.~\textbf{96}, 113002 (2006).

\bibitem{jacob}
Z.~Jacob, I.~I.~Smolyaninov and E.~E.~Narimanov,
Broadband Purcell effect: radiative decay engineering with metamaterials,
Appl.~Phys.~Lett.~\textbf{100}, 181105 (2012).

\bibitem{krasnok}
A.~E.~Krasnok, A.~P.~Slobozhanyuk, C.~R.~Simovski, S.~A.~Tretyakov, A.~N.~Poddubny,
A~.E.~Miroshnichenko, Y.~S.~Kivshar and P.~A.~Belov,
An antenna model for the Purcell effect,
Sci.~Rep.~\textbf{5}, 12956 (2015).

\bibitem{balanis}
C.~A.~Balanis, \emph{Antenna Theory, Analysis and Design} (John Wiley \& Sons, NY, 2005).


\end{thebibliography}
\end{document}